\documentclass{article}
\usepackage[vmargin=1in,hmargin=0.75in]{geometry}
\usepackage{amsmath}
\usepackage{amssymb}
\usepackage{amsfonts}
\usepackage{graphicx}
\usepackage{url}

\usepackage{graphics}
\usepackage{epsfig}
\usepackage{tikz}
\usepackage{subfig}
\usepgflibrary{shapes.geometric}
\usepackage[T1]{fontenc}

\newcommand{\Si}{\Sigma}

\usepackage{verbatim}

\newcommand{\Ber}{\mbox{Ber}}

\begin{document}

\title{Model Selection Approach for Distributed Fault Detection in Wireless Sensor Networks}

\author{Mrinal Nandi $^{\rm a}$\thanks{Corresponding author. Email: mrinal.nandi1@gmail.com}, Anup Dewanji $^{\rm b}$, Bimal Roy $^{\rm b}$ and Santanu Sarkar $^{\rm c}$\\
$^{\rm a}${\em{Department of Statistics, West Bengal State University, Barasat, India}};\\ 
$^{\rm b}${\em{ASD, Indian Statistical Institute, 203 B. T. Road, Kolkata-700 108, India}};\\
$^{\rm c}${\em{Chennai Mathematical Institute, Chennai, India.}} }

\maketitle

\begin{abstract} 
Sensor networks aim at monitoring their surroundings for event detection and object tracking. But, due to failure, 
or death of sensors, false signal can be transmitted. In this paper, we consider the problems of distributed fault 
detection in wireless sensor network (WSN). In particular, we consider how to take decision regarding fault detection 
in a noisy environment as a result of false detection or false response of event by some sensors, where the sensors are 
placed at the center of regular hexagons and the event can occur at only one hexagon. We propose fault detection 
schemes that explicitly introduce the error probabilities into the optimal event detection process. We introduce 
two types of detection probabilities, one for the center node, where the event occurs and the other one for the adjacent nodes. 
This second type of detection probability is new in sensor network literature. We develop schemes under the model selection 
procedure, multiple model selection procedure and use the concept of Bayesian model averaging to identify a 
set of likely fault sensors and obtain an average predictive error.
\end{abstract}

 
\section{Introduction}
Traditional and existing {\it sensor-actuator networks} use wired communication, whereas wireless sensor networks 
(WSN) provide radically new communication and networking paradigms and myriad new applications. The wireless sensors have small size, 
low battery capacity, non-renewable power supply, small processing power, limited buffer capacity and low-power radio. They may measure
distance, direction, speed, humidity, wind speed, soil makeup, temperature, chemicals, light, and various other parameters.  

Recent advancements in wireless communications and electronics have enabled the development of low-cost WSN. A WSN usually consists of 
a large number of small sensor nodes, which are equipped with one or more sensors, some processing circuit, and a wireless transceiver.
One of the unique features of a WSN is random deployment in inaccessible terrains and cooperative effort that offers unprecedented 
opportunities for a broad spectrum of civilian and military applications, such as industrial automation, military surveillance, national 
security, and emergency health care~\cite{ASSC,PK,A}. Sensor Networks are also useful in detecting topological events such as forest fires~\cite{FSAW}.

Sensor networks aim at monitoring their surroundings for event detection and object tracking~\cite{ASSC,MS}. Because of this surveillance 
goal, {\it coverage} is the functional basis of any sensor network. In order to fulfill its designated tasks, a sensor network must 
fully cover the Region of Interest (ROI) without leaving any {\it internal sensing hole}~\cite{B,BS,C,FKP}. So far, a number of 
movement-assisted sensor placement algorithms have been proposed. An exclusive survey on these topics is presented by Li et al.~\cite{L}. 
On the other hand sensor could die or fail at runtime for various reasons such as power depletion, hardware defects, etc. So, even after 
the ROI is fully covered by the sensors, wrong information can be communicated by some sensors, or sensors may fail to detect the event 
due to noise or obstructions. Chen et al.~\cite{CKS} have proposed a distributed localized fault detection algorithm for WSN, where each 
sensor identifies its own status to be either good or faulty and the claim is then supported or reverted by its neighbors. The proposed 
algorithm is analyzed using a probabilistic approach. Sharma et al.~\cite{S} have characterized the different types of fault and proposed 
a different algorithm for fault detection considering different types of fault. Some of the methods are statistical, like, using histogram, 
etc. Both the work can only detect the faulty sensors, but not the event.

One of the important sensor network applications is monitoring inaccessible environments. Sensor networks are used to determine event 
regions and boundaries in the environment with a distinguishable characteristic~\cite{KI,CG,NM}. The basic idea of distributed 
detection~\cite{T} is to have each of the independent sensors make a local decision (typically, a binary one, i.e., an event occurs or 
not) and then combine these decisions at a fusion sensor (the sensor which collects the local information and takes the decision), or 
at a base station to generate a global decision. 

\subsection{Our Motivation}
In this paper, we are interested in one particular query: determining event in the environment (i.e., ROI) with a distinguishable 
characteristic. We assume the ROI to be partitioned into suitable number of congruent regular hexagonal cells (i.e., we can think ROI 
as a regular hexagonal grid). This physical structure of ROI is not a requirement for the theoretical analysis, we can do the similar analysis
with other structure also. Suppose that sensors are placed a priori at the center (which are known as nodes) of every hexagon of the grid. We assume that 
the sensors are connected to its adjacent sensor nodes in the sense that a hexagon will be strongly covered by its center node and weakly
covered by the adjacent nodes. If event occurs in the hexagon where a particular sensor lies, then that particular sensor can detect the
event with a greater probability whereas, if event occurs in any adjacent hexagon, then the particular sensor can detect the event with 
a lesser probability. Hence, only one node (center node of the event hexagon) can detect the event hexagon with greater probability, 
say $p_1$, and adjacent nodes (six for interior nodes and less for boundary nodes) can detect the event hexagon with lesser probability, 
say $p_2$, with $p_1 > p_2$. We assume that no other sensor can detect the event hexagon. In this paper, unlike the previous works, we 
assume that if the event occurs then it occurs at only one hexagon of the grid which will be known as event hexagon and there is no 
fusion sensor. All sensors can communicate with the base station and the base station takes the decision about the query. As an example, 
consider a network of devices that are capable of sensing mines or bombs, if we assume that a few mines or bombs may be placed on a 
particular area of ROI. Information from these devices can be sent to a nearby police station, or a central facility. Then, an important 
query in this situation could be whether a particular hexagon is the event hexagon or not (i.e., mines or bombs are placed or not). 

One fundamental challenge in the event detection problem for a sensor network is the detection accuracy which is disturbed by the noise 
associated with the detection and the reliability of sensor nodes. A sensor may fail to detect the event due to natural obstruction or 
any other causes. After detecting the event, a sensor can send false message to the base station due to some technical reasons. The 
sensors are usually low-end inexpensive devices and sometimes exhibit unreliable behavior. For example, a faulty sensor node may issue 
an alarm even though it has not received any signal for event, or it cannot detect any event, and vice versa. Moreover, a sensor may be 
dead, in which case, the sensor cannot send any alarm.  

\subsection{Our Contribution}
In our theoretical analysis, the sensor fault probabilities are introduced into the optimal event detection process. We apply model 
selection approach, multiple model selection approach and Bayesian model averaging methods~\cite{HMRV,MY} to find a solution of the 
problem. We develop the schemes using the model selection technique. We calculate different error probabilities and find some theoretical results. 

In all previous works, the authors assume only one detection probability. We introduce two detection probabilities, $p_1$ and $p_2$, one 
for the center node and other for the adjacent nodes. Even if the center node may fail to detect the event, the adjacent nodes may 
detect the event, and vice versa. We consider these probabilities and show that, in various situations, the adjacent nodes play key role 
to detect the event. One can introduce more detection probabilities and analyze the situation in similar manner.
  
The parameters $p_1$ and $p_2$, the detection probabilities of a sensor, and error probabilities (see Section \ref{Analysis}) cannot be 
estimated from the real life situations, but need to be estimated beforehand by some experimentation. The prior probabilities of various 
events also cannot be estimated, but may be known in some cases. Finally, we calculate the error probabilities numerically for some 
values of the parameters of our model and make some concluding remarks analyzing the results. 
 
\section{Previous Work}
Lou et al.~\cite{LDH} consider two important problems for distributed fault detection in WSN: 1) how to address both the noise-related 
measurement error and sensor fault simultaneously in fault detection and 2) how to choose a proper neighborhood size $n$ for a sensor 
node in fault correction such that the energy could be conserved. They propose a fault detection scheme that explicitly introduces the 
sensor fault probability into the optimal event detection process. They show that the optimal detection error decreases exponentially 
with the increase of the neighborhood size.  

Krishnamachari and Iyengar~\cite{KI} propose a distributed solution for canonical task in WSN (i.e., the binary detection of interesting 
environmental events). They explicitly take into account the possibility of sensor measurement faults and develop a distributed Bayesian 
algorithm for detecting and correcting such faults. 

Nandi et al.~\cite{MBA} consider the problem of distributed fault detection in wireless sensor network (WSN), where the sensors are 
placed at the center of a particular square (or hexagon) of the grid covering the ROI. They proposed fault detection schemes that 
explicitly introduce the error probabilities into the optimal event detection process. They developed the schemes under the consideration 
of Neyman-Pearson hypothesis test and Bayes test. They also calculate type I and type II errors for different values of the parameters.

In almost all the previous works, except~\cite{MBA}, authors assume that event occurs over a region and there are fusion sensors that 
collect the information locally and take a decision. Since they do not introduce the concept of base station there is no concept of 
response probability. Also, they assume informations are spatially correlated. Unlike the previous work, in this paper, we assume that 
if event occurs then it occurs at only one cell of the ROI and there is no fusion sensor. All the sensors send information to the base 
station. We introduce the probability model in two different stages; firstly, when a sensor detects the event and, secondly, when a 
sensor sends the message to the base station. In the previous works, only one type of detection probability has been introduced to 
simulate the different error probabilities for some specific values of the parameters. In this paper, we introduce two different 
detection probabilities and obtain analytically the exact test and estimate the error probabilities by simulation. In almost all the 
previous works, authors assume the ROI to be a square grid. The hexagonal grid is better in the sense that minimum number of sensors 
are required to cover the entire ROI~\cite{W}.

\section{Statement of the problem and Assumptions}
In this section, we describe the problem in more specific terms and state the assumptions that we make.

\begin{figure}
\begin{center}
\caption{Nodes placed  in centers when ROI partitioned in to regular hexagons}
\resizebox{8 cm}{8 cm}{

\begin{tikzpicture}[scale=1.87]
\label{hexagon}
\node [draw, thick, minimum size=4.2cm, regular polygon, regular polygon sides=6] at (2,2) {center \ node};
\node [draw, thick, minimum size=3cm, regular polygon, regular polygon sides=6] at (2,3.92) {adjacent \ node};
\node [draw, thick, minimum size=3cm, regular polygon, regular polygon sides=6] at (2,0.08) {adjacent \ node};
\node [draw, thick, minimum size=3cm, regular polygon, regular polygon sides=6] at (0.36,2.96) {adjacent \ node};
\node [draw, thick, minimum size=3cm, regular polygon, regular polygon sides=6] at (0.36,1.04) {adjacent \ node};
\node [draw, thick, minimum size=3cm, regular polygon, regular polygon sides=6] at (3.64,2.96) {adjacent \ node}; 
\node [draw, thick, minimum size=3cm, regular polygon, regular polygon sides=6] at (3.64,1.04) {adjacent \ node};
\end{tikzpicture}
}
\end{center}
\end{figure}
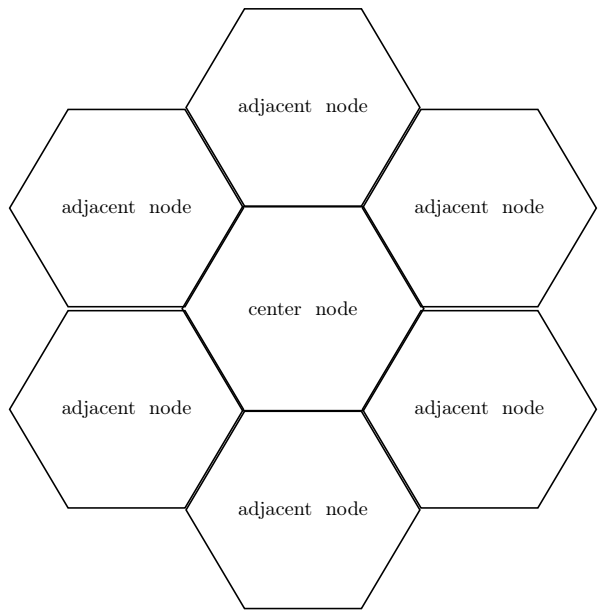

Sensors are deployed, or manually placed, over ROI to perform event detection (i.e., to detect whether an event of interest has 
happened or not) in ROI. If sensors are deployed from air then, using actuator-assisted sensor placement or by movement-assisted sensor 
placement, sensors are so placed that sensor network covers the entire ROI. This ROI is partitioned into suitable number of regular 
hexagons (i.e., we can think of the ROI as a regular hexagonal grid), as shown in Figure~\ref{hexagon}. Sensors are placed a priori at 
every center (which are known as nodes) of the regular hexagons. Sensors have two detection probabilities. The sensor network covers the
entire ROI and there is only one event hexagon, as discussed before.

Each sensor node determines its location through beacon positioning mechanisms ~\cite{BHE} or by exploiting the Global Positioning 
System (GPS). Through a broadcast or acknowledge protocol, each sensor node is also able to locate the neighbors within its communication 
radius. Sensors are also able to communicate with the base station. Base station will take the decision. In this paper, we assume that, 
event occurs at one particular hexagon of the grid which will be known as {\it event hexagon} or event does not occur (in that case we 
say ROI is {\it normal}). All sensors can communicate with the base station and base station takes the decision by combining the 
information received from all the sensors. 

There are two phases in the whole process. The first one is detection phase, when the sensor at the center of a regular hexagon tries 
to detect the event. The sensor at the center of the event hexagon can detect the event hexagon with greater probability $p_1$ and the 
sensors at the  adjacent nodes (see Figure~\ref{hexagon}) can detect the event hexagon with lesser probability $p_2$. We also assume 
that there is a prior probability that a particular hexagon is an event hexagon. The next phase is response phase, in which sensors 
send message to the base station. Even if the event hexagon is detected by a sensor, it may not respond (i.e., send message to the base 
station that no event occurred in that cell and the neighboring cells due to some technical fault) with some probability; then we say 
that the sensor is a {\it faulty sensor}. Conversely, if event hexagon is not detected, or there is no event hexagon at all (i.e., ROI 
is normal), then also a faulty sensor can send the wrong information to the base station with some probability. A sensor is said to be 
a {\it dead sensor} if the sensor does not work. A dead sensor sends no response in either cases.

Each sensor sends information to the base station. As the sensors may send wrong information, the base station takes the important role 
in identifying the event hexagon. Base station will collect all the information and take a decision about the event hexagon according 
to a rule which we have to find out. Our job is to find a rule for the base station such that base station works most efficiently.

\subsection{Notations and Assumption} 
Our problem is to develop a strategy for the base station to take decision about event hexagon (i.e., which hexagon of the ROI is the 
event hexagon, if at all). Let $R$ be the set of all nodes.
% and $R'$ be the set of all interior nodes (node which has six adjacent nodes). 
For $N \in R$, define $B(N)$, as the set of adjacent node(s) of $N$ and $k(N)$ be the number of adjacent node(s) of $N$. Hence, 
$0\leq k(N) \leq 6$. Call a node $N$ interior if $k(N)=6$. Let $S_N$ be the sensor which is placed at the node $N$ and $H_N$ be the
hexagon where the node $N$ is placed (i.e., $N$ is the center of $H_N$). For $N \in R$, let $X_N$ denote the true status of the node 
$N$. That is, $X_N=1$ if event occurs at $H_N$, and $0$ otherwise. Also define $Y_N=0$ if $S_N$ detects no event, and $1$ if $S_N$ 
detects the event in $H_N$ or $H_{N'}$, for $N' \in B(N)$. Finally define $Z_N=0$ if $S_N$ does not respond, i.e., the sensor informs 
the base station that event does not occur at $H_N$ or $H_{N'}$ for $N' \in B(N)$, and $Z_N=1$ if $S_N$ responds, i.e., the sensor 
$S_N$ informs the base station that the event has occurred in $H_N$ or $H_{N'}$, for $N' \in B(N)$.

Now we make one natural assumption that, once detection phase is completed, response of a sensor depends only on what it detects but 
not on whether the event has actually occurred or not, i.e., $P(Z_N=k|Y_N,X_N)= P(Z_N=k|Y_N), \mbox{\ for \ } k=0,1.$ We also assume 
that the sensors work independently and identically.

Since we assume that there is at most one event hexagon, $\sum_{N\in R} X_N = 1$ or $0$.\\
The possible true scenarios are, therefore, represented by the following $|R|+1$ different models:

${\cal M}_0$ : ($X_N = 0$ for all $N \in R $), 

and, for each $N \in R $,

${\cal M}_N$ : ($X_N = 1$ and $X_{N'} = 0$ for all $N' \in R \setminus{N}$).
    
$\mbox{Let \ } \Pr({\cal M}_0) =  P(\mbox{ROI is normal}) = p_{norm}$\\
 and, for all $N \in R$, $ \Pr({\cal M}_N) = \Pr(\mbox {event occurs at the hexagon\ } H_N) = p_N.$\\  
In particular, we may assume $p_N$'s to be same for all $N$. 
We denote any probability under the model ${\cal M}_0$ as $P_{{\cal M}_0}(\cdot)$ and under the model ${\cal M}_N$ as $P_{{\cal M}_N}(\cdot)$.

We also make the followings assumptions:
\begin{itemize}
\item For all $N \in R$, $P_{{\cal M}_0}(Y_N=1) = 0$ and $P_{{\cal M}_N}(Y_N=1) = p_1.$
\item For all $N' \in B(N), P_{{\cal M}_N}(Y_{N'}=1) = p_2$, and \\
$ \mbox{for all\ } N' \in R \setminus [B(N)\cup \{N\}],  P_{{\cal M}_N}(Y_{N'}=1) = 0 .$
\item For all  $N \in R, P(Z_N=1|Y_N=1) = p_c$ and $P(Z_N=1|Y_N=0) = p_w $.
\item $Z_N$ and $Y_{N'}$ are independent for $N \neq N'$. 
\item The responses from different nodes are independent under a particular model, i.e., $Z_N$'s are independent under ${{\cal M}_{N'}}$ for a fixed $N'\in R$.
\end{itemize}

\section{Theoretical Analysis of fault Detection} 
\label{Analysis}
In this section we discuss some theoretical results. In real situations, $|R|$ may be very large. Given the network of the sensor nodes 
and some prior knowledge about the nature of event, one may have fairly good idea about the set of feasible regions for the event. 
Formally, instead of all possible models, one may be able to restrict to a set containing all the feasible models. For example, if the 
event is known to take place in a particular region, we can restrict our models accordingly. 

\subsection{Model Selection Approach}
\label{Model}
\begin{eqnarray*}
\mbox{For all } N\in R, P_{{\cal M}_0}(Z_N=1)\hspace*{230pt}\\
 =P_{{\cal M}_0}(Z_N=1|Y_N=0) P_{{\cal M}_0}(Y_N=0)
   +P_{{\cal M}_0}(Z_N=1|Y_N=1) P_{{\cal M}_0}(Y_N=1)\\
 = P(Z_N=1|Y_N=0) P_{{\cal M}_0}(Y_N=0)
   + P(Z_N=1|Y_N=1)P_{{\cal M}_0}(Y_N=1) = p_w.
\end{eqnarray*}
Hence, under the model ${\cal M}_0, Z_N$ follows $\Ber(p_w)$, for all $N\in R$,
and the likelihood of the data \{$Z_N=z_N,\mbox{\ for all \ } N\in R$ \}, under the model ${\cal M}_0$, is 
\begin{eqnarray*}
 L_0 = P_{{\cal M}_0}(Z_N=z_N, \mbox{\ for all\ } N \in R )\hspace*{150pt}\\
 \hspace*{50pt}=\prod_{N\in R}p_w^{z_N}(1-p_w)^{(1-z_N)} = (p_w)^{\sum_{N\in R}{z_N}}\times (1-p_w)^{\sum_{N\in R} \ {(1-z_N)}}.                                                   & 
\end{eqnarray*}
So  $\ln L_0 = \Si_{N\in R}\ {z_N}\ln p_w + \Si_{N\in R}\ {(1-z_N)}\ln(1-p_w).$
%$$ \mbox{For any}~N \in R,~\mbox{we have}~ P_{{\cal M}_N}(Z_N=1), $$
\begin{eqnarray*}
  &&\mbox{For any}~N \in R,~\mbox{we have}~ P_{{\cal M}_N}(Z_N=1)\\
  &=&P_{{\cal M}_N}(Z_N=1|Y_N=0) P_{{\cal M}_N}(Y_N=0) + P_{{\cal M}_N}(Z_N=1|Y_N=1) P_{{\cal M}_N}(Y_N=1)\\
  &=&P(Z_N=1|Y_N=0) P_{{\cal M}_N}(Y_N=0) + P(Z_N=1|Y_N=1)P_{{\cal M}_N}(Y_N=1)\\
  &=&p_w(1-p_1)+ p_cp_1 = p_1(p_c-p_w)+ p_w = P_1,~\mbox {say.}
\end{eqnarray*}
Hence, for all $N \in R$, under ${\cal M}_N, Z_N$ follows $\Ber(P_1)$. Similarly, for all $N' \in B(N)$, under ${\cal M}_N, Z_{N'}$ follows $\Ber(P_2)$, 
where $P_2= p_2(p_c-p_w)+ p_w $ and, under ${\cal M}_N, Z_{N'}$ follows Ber$(p_w)$ for all $N' \in R\setminus [B(N)\cup \{N \}]$. Note that $P_1>P_2$ since $p_1>p_2$. 
Hence the likelihood for the model ${\cal M}_N$, given $Z_{N'}=z_{N'}, N'\in R$, is 
\begin{eqnarray*}
&&L_N=P_{{\cal M}_N}(Z_{N'}=z_{N'}, \mbox{\ for all\ } N' \in R )=\\
&&P_1^{z_N}(1-P_1)^{(1-z_N)}\Pi_{N'\in B(N)} P_2^{z_{N'}}(1-P_2)^{(1-z_{N'})}\\
&&\times \Pi_{N'\in R\setminus [B(N)\cup \{N\}]} \ p_w^{z_{N'}}(1-p_w)^{(1-z_{N'})}\\
&&= P_1^{{z_N}}(1-P_1)^{(1-z_N)}P_2^{\Si_{N'\in B(N)}{z_{N'}}}(1-P_2)^{\Si_{N'\in B(N)}{(1-z_{N'})}}\\
&&\times p_w^{\Si_{N'\in R\setminus [B(N)\cup \{N\}]}{z_{N'}}}(1-p_w)^{\Si_{N'\in R\setminus [B(N)\cup \{N\}]}{(1-z_{N'})}}.\\
&&\mbox{Let \ }T_N=\sum_{N'\in B(N)}{Z_{N'}}, \mbox{ \ \ so that} \sum_{N'\in B(N)}{(1-Z_{N'})}=k(N)-T_N\\
&&\mbox{with the corresponding observed values denoted by} \\
&&t_N=\sum_{N'\in B(N)}{z_{N'}}  \mbox{ \ and \ }  \sum_{N'\in B(N)}{(1-z_{N'})}=k(N)-t_N.\\
&& \mbox {Therefore,\ }\ln L_N = \\
&&z_N\ln P_1 + (1-z_N)\ln(1-P_1) + t_N\ln P_2 +(k(N)-t_N)\ln (1-P_2)\\
&&+ \sum_{N'\in R\setminus [B(N)\cup \{N\}]} \ z_N\ln p_w + \sum_{N'\in R\setminus [B(N)\cup \{N\}]} \ (1-z_N)\ln(p_w(1-p_w))\\
&&= \ln L_0 + z_N \ln \frac{P_1}{p_w} + (1-z_N)\ln \frac{1-P_1}{1-p_w} + t_N\ln \frac{P_2}{p_w} + (k(N)-t_N)\ln \frac{1-P_2}{1-p_w}\\
&&= \ln L_0 + z_N \ln \frac{P_1(1-p_w)}{p_w(1-P_1)} + t_N \ln \frac{P_2(1-p_w)}{p_w(1-P_2)} + \ln \frac{1-P_1}{1-p_w} + k(N)\ln \frac{1-P_2}{1-p_w}\\
&&= a+ b(cz_N + t_N - dk(N)), \mbox{ say,}\\
&&\mbox{where, \ } a=\ln L_0 + \ln \frac{1-P_1}{1-p_w}, b=\ln \frac{P_2(1-p_w)}{p_w(1-P_2)}> 0, \\
&&c= \frac{\ln \frac{P_1(1-p_w)}{p_w(1-P_1)}}{\ln \frac{P_2(1-p_w)}{p_w(1-P_2)}} \mbox{ \ and\ } d= \frac{\ln \frac{1-p_w}{1-P_2}}{\ln \frac{P_2(1-p_w)}{p_w(1-P_2)}}\mbox{ are independent of } N.\\
\end{eqnarray*}
In model selection approach, the model resulting in the maximum value of the likelihood is selected. Note that, since there is no 
parameter being estimated, this is equivalent to the well-known Akaike Information Criterion(AIC)~\cite{GDS}. Therefore, the base 
station will accept the model ${\cal M}_0$ if 
\begin{eqnarray*}
 &=& \ln \frac{1-P_1}{1-p_w} + b(cz_N + t_N - dk(N)) <0, \mbox{for all \ } N \in R. 
\end{eqnarray*}
Otherwise, as $b$ is positive, accept the model ${\cal M}_N$ for which $(cz_N + t_N - dk(N))$ is maximum among all \ $N\in R $. If 
values of $(cz_N + t_N - dk(N))$ are equal for more than one $N$, then we can select one of the corresponding models with equal 
probability. If we want to maximize the likelihood for the models ${\cal M}_N$ corresponding to the interior nodes only, so that 
$k(N)$ is fixed, then we need to maximize $(cz_N + t_N)$ among all $N \in R$.

\subsection {Multiple Model Selection}
\label{Multi}
Instead of selecting one particular model, one may want to select more than one models with approximately similar log likelihood values 
to the maximum one. We can consider the set of models 
\begin{eqnarray*}
\{{\cal M}_K: \frac{L_K}{max_{N \in R}L_N} > C \},
\end{eqnarray*}
where $0<C<1$ is a suitable constant close to $1$. This $C$ is usually chosen according to the resource available. This is similar to the idea of Occam's window in the context of Bayesian model selection~\cite{HMRV}. This may be interpreted as the interval estimation for the true model.

Note that $L_N$ is an increasing function of $cz_N + t_N-dk(N)$, as $b$ is positive. 
We consider only the following set of models
\begin{eqnarray*}
 \{{\cal M}_K: Q_K > C^*\cdot max_{N\in R}\  Q_N \},
\end{eqnarray*}
where $Q_N=cz_N + t_N-dk(N),$ for all $N \in R,$ with $0<C^*<1.$
In particular, if we consider the interior nodes only, then we consider the set of models given by
$$ \{{\cal M}_K: cz_K + t_K> C^*\cdot max_{N\in R}\{cz_N + t_N \} \}.$$
We can select multiple models using some other criteria. One such may be to select all the models (one or more) for which the maximum value of the likelihood is attained. 
Let ${\cal N}_{\max}$ be the set of nodes corresponding to all these models, including $`N=0$' corresponding to ${\cal M}_0$ if it has the 
maximum value of the likelihood. Then this method select all the models ${\cal M}_N$ with $ N\in {\cal N}_{\max}$.
By another criterion, one may select the models ${\cal M}_{N'}$, for $ N' \in {\cal N}_{\max}\cup[\cup_{N\in {\cal N}_{\max}} B(N)]$; that is, $N'$ be a node in ${\cal N}_{\max}$ or any of the neighboring nodes of a node in ${\cal N}_{\max}$. 
Note that $B(N)$ for $N=0$ is the empty set. One can combine these two types of criteria and come up with many others.
\subsection{Bayesian Model Averaging}
Bayesian model averaging is an effective method to solve a decision problem when there are many alternative hypotheses or models, which are complicated~\cite{HMRV}. Suppose ${\cal M}_1, {\cal M}_2,\ldots,{\cal M}_k$ are the models considered and $D$ denotes the given data. The posterior probability for model ${\cal M}_k$ is given by 
$$\Pr({\cal M}_k|D)= \frac{\Pr(D|{\cal M}_k)\Pr({\cal M}_k)}{\sum\Pr(D|{\cal M}_l)\Pr({\cal M}_l)},$$
where $\Pr(D|{\cal M}_k)$ denotes the probability of observing data $D$ under the model ${\cal M}_k$ (which is essentially the likelihood $L_k$ under ${\cal M}_k$) and $\Pr({\cal M}_k)$ is the prior probability that ${\cal M}_k$ is the true model (assuming that one of the models is true). 

In this work, the data $D$ is $\{Z_N= z_N : N\in R \}$ and the models are ${\cal M}_0, {\cal M}_N, N \in R$ as defined in Section 3.2. Hence, The posterior probability for model ${\cal M}_N$ is 
$$\Pr({\cal M}_N|Z_N=z_N, N\in R)= \frac{p_NL_N}{\sum_{l \in R} p_lL_l+p_{norm} L_0},$$ 
$$\mbox{and that for \ }{\cal M}_0 \mbox{\ is \ } \frac{p_{norm}L_0}{\sum_{l \in R} p_lL_l+p_{norm} L_0}.$$ 

We select the model ${\cal M}_0$ if $p_{norm}L_0$ is greater than $p_NL_N$, for all $N \in R$; otherwise, select ${\cal M}_N$ for which $p_NL_N$ is maximum among all $N \in R$. Hence, if $p_N$'s are all equal, then Bayesian approach is same as the likelihood approach.
\section{Some Important Considerations and Error Probabilities}
In this section, we consider some important issues related to the problem of fault detection and the proposed methodology including calculation of errors (e.g., false detection, etc.) and detection probabilities.

The following probabilities give some idea about the role of neighboring nodes, along with the center node, in detection, or false detection, 
of event. For example, $P_{{\cal M}_0}(T_N=0, Z_N=1)$ gives the probability of a false detection by the $N^{th}$ node, and not by the neighboring nodes, 
while $P_{{\cal M}_N}(T_N=6,Z_N=0)$ gives the probability of a false negative by the $N^{th}$ node, with all the neighboring nodes detecting the event. Since, given a particular model, $T_N$ and $Z_N$ are independent, calculation of such probabilities is simple as given in the following. 
For any $ N\in R$ and $i=0,1,\ldots,k(N)$, 

\begin{enumerate}
\item $ P_{{\cal M}_0}(T_N=i, Z_N=0)={k(N)\choose i}p_w^i (1-p_w)^{k(N)-i+1}$
\item $ P_{{\cal M}_0}(T_N=i, Z_N=1) = {k(N)\choose i}p_w^{i+1}(1-p_w)^{k(N)-i}$
\item $P_{{\cal M}_N}(T_N=i, Z_N=0) =	{k(N)\choose i} P_2^i(1-P_2)^{k(N)-i}(1-P_1)$
\item $P_{{\cal M}_N}(T_N=i, Z_N=1) = {k(N)\choose i} P_2^i(1-P_2)^{k(N)-i}P_1$.
\end{enumerate}
\begin{eqnarray*}
&&\mbox{Note that, for}~ N\in R,P_{{\cal M}_0}(L_N> L_0)=P_{{\cal M}_0}(\ln L_N> \ln L_0)=\\
&&P_{{\cal M}_0} \left(Z_N \ln \frac{P_1(1-p_w)}{p_w(1-P_1)}+ T_N \ln \frac{P_2(1-p_w)}{p_w(1-P_2)}+ \ln \frac{1-P_1}{1-p_w}+  k(N) \ln \frac{1-P_2}{1-p_w} > 0 \right)\\
&&=P_{{\cal M}_0}\left(Z_N \ln \frac{P_1(1-p_w)}{p_w(1-P_1)}+ T_N \ln \frac{P_2(1-p_w)}{p_w(1-P_2)}> k(N) \ln \frac{1-p_w}{1-P_2} + \ln \frac{1-p_w}{1-P_1}\right),\\
\end{eqnarray*}
which can be numerically obtained using the joint distribution of $T_N$ and $Z_N$ under the model ${\cal M}_0$.
The maximum of these probabilities over all $N$ gives a lower bound for the probability that a node is considered to be an event node when the ROI is normal. 
On the other hand, the sum over all $N$ gives an upper bound for the same. 
Similarly, for $N\in R$,  $P_{{\cal M}_N}(L_N < L_0)= $
$$P_{{\cal M}_N}\left(Z_N \ln \frac{P_1(1-p_w)}{p_w(1-P_1)}+ T_N \ln \frac{P_2(1-p_w)}{p_w(1-P_2)} < k(N) \ln \frac{1-p_w}{1-P_2} + \ln \frac{1-p_w}{1-P_1}\right),$$
which can be again numerically obtained using the joint distribution of $T_N$ and $Z_N$ under the model ${\cal M}_N$. This probability gives some idea about the error that, when $N^{th}$ node is the event node and it is not detected.
  
% For $N,N'\in R, \ N\neq N'$,  $P_{{\cal M}_N}(L_{N'}> L_N)$
%$$=P_{{\cal M}_N}((Z_{N'}-Z_N) ln \frac{P_1(1-p_w)}{p_w(1-P_1)}+ $$
%$$(T_{N'}-T_N) ln \frac{P_2(1-p_w)}{p_w(1-P_2)}> (k(N')-k(N)) ln \frac{1-p_w}{1-P_2}.$$
As noted in Section \ref{Model}, we select the model ${\cal M}_N$ for which $Q_N$ is the maximum, for $N \in R$. The random variable $Q_N$ is, therefore, of some interest, the distribution of which under different models is useful in calculating many error probabilities. We first find the distribution of $Q_N$ under the model ${\cal M}_N$. Note that $Q_N$ takes values $ci+j-dk(N)$, corresponding to $Z_N=i$ and $T_N=j$, for $i=0,1$, and $j=0,1,2,\ldots,k(N)$. Assume that, for convenience, the values of $Q_N$ for different $i$ and $j$ are all distinct. 
Therefore, for $i=0,1$ and $j=0,1,\ldots,k(N)$,
\begin{eqnarray*}
&& P_{{\cal M}_N} \left( Q_N=ci+j-dk(N) \right)=   {k(N)\choose j}(P_1)^i (1-P_1)^{(1-i)}(P_2)^j (1-P_2)^{(k(N)-j)} \\
&& \mbox{and, \ }P_{{\cal M}_0} \left(Q_N=ci+j-dk(N) \right)= {k(N)\choose j}(p_w)^{i+j} (1-p_w)^{(1-i+k(N)-j)}. \\
\end{eqnarray*}
For $N' \in B(N)$, or $N' \in R\setminus [B(N)\cup \{N\}]$, one can find $P_{{\cal M}_{N'}}(Q_N=ci+j-dk(N))$ in similar manner, although the calculation 
is very tedious as there are many sub-cases. Ideally, one is interested in probability of errors occurring at the level of base station. 
For example, the two important errors are: (1) not selecting ${\cal M}_0$ when ${\cal M}_0$ is true (false positive), and (2) selecting ${\cal M}_0$ when ${\cal M}_N$ is true for some $N\in R$ (false negative). 
Theoretical calculation of these error probabilities is complicated. We, therefore, use simulation technique to estimate these and similar error probabilities.

\section{Simulation Study}
We consider a 32 $ \times $ 32 hexagonal grid and we run the programme 10000 times. The simulation is performed using the C-code, and required random numbers are generated using the standard C-library.

In our simulation study, we consider different criteria, as discussed in Sections \ref{Model} and \ref{Multi}, for estimating the error probabilities, or equivalently, the success rate.
First consider the probability of selecting ${\cal M}_0$, when it is true. Let $S_1$ denote the proportion of correct detection of normal situation, when model $ {\cal M}_0$ is true, using the model selection method of Section 4.1. That is, $S_1$ gives an estimate of $P_{{\cal M}_0}( 0\in {\cal N}_{max} \mbox{~and $0$ is selected by randomization})$. Then $1-S_1$ gives an estimate of the false positive rate.

When ${\cal M}_N$ is true for some $N \in R$, let $S_2$ denote the proportion of correct decision for the event node using the model selection method of Section \ref{Model}, 
so that it estimates $P_{{\cal M}_N}(N \in {\cal N}_{\max} \mbox{~and is selected by randomization}).$ Note that, for each simulation run, the event hexagon is chosen randomly so that $S_2$ gives an average value over all $N$. 
In this context, this probability is same for all the interior nodes. Then, $1-S_2$ gives an estimate of the corresponding error probability of not selecting ${\cal M}_N$, when it is true.

Note that, in this problem of fault detection with a single event node, the likelihood value, for a given observed data configuration, may be equal for more than one models. 
Therefore, quite often, the maximum value of the likelihood may be attained by more than one model. The model selection method of Section \ref{Model}, which selects one of these models randomly in such cases, may often not select the correct model.
Therefore, the method of Section \ref{Multi}, which selects more than one models having similar likelihood value, may be preferred and will have better chance of selecting the correct model. We now consider some of those methods in the following. 

Let us first consider the method in which all the models corresponding to the maximum value of the likelihood are selected. Let $S_3$ denote the proportion of correct selection of the model ${\cal M}_N$, when it is true, by this method.
Then $S_3$ estimates the probability $P_{{\cal M}_N}(N \in {\cal N}_{\max})$, which is always more than or equal to the quantity estimated by $S_2$, as remarked before.
We also consider the method in which all the models having maximum likelihood along with their neighborhood models are selected. 
A model ${\cal M}_{N'}$ is a neighborhood model of the model ${\cal M}_N$ if $N'$ is a neighboring node of $N$. If $S_4$ denotes the proportion of correct selection of the model ${\cal M}_N$, when it is true, by this method, then $S_4$ estimates $ P_{{\cal M}_N}(N \in {\cal N}_{\max}\cup\{\cup_{N'\in {\cal N}_{\max}} B(N')\})$. 
Clearly, $S_4\geq S_3\geq S_2$. Similarly, if $S_5$ denotes the proportion of correct selection of the model ${\cal M}_N$, when it is true, by selecting all those models with likelihood value being more than $90\%$ of the maximum likelihood (that is, the method of Section \ref{Multi} with $C=0.9$) then $S_5$ estimates the probability $P_{{\cal M}_N}(L_N>0.9 L_{max})$ with $L_{max}$ denoting the maximum value of the likelihood.

\begin{table}[htb]
\scriptsize
\caption{Simulation of estimated probabilities for some values of the parameters}
{
\begin{center}
\begin{tabular}{|c|c|c|c|c|c|c|c|c|c|c|c|c|c}
\hline
\multicolumn{3}{|c|}{other parameters} & \multicolumn{8}{|c|}{Simulation of different probabilities with $p_c= 0.9$} \\ \hline \hline
    $p_1$& $p_2$& $p_w$& $S_1$ & $S_2$    &$S_3$ & $N_3$& $S_4$ & $N_4$& $S_5$& $N_5$  \\ \hline
     0.9 &  0.0 & 0.01  & 0.00 &  0.08    & 0.81  & 18.16 & 0.81  & 59.85 & 0.82 & 18.17 \\ \hline
     0.9 &  0.3 & 0.01  & 0.00 &  0.47    & 0.69  & 5.44 & 0.70  & 14.81 & 0.73 & 5.70 \\ \hline 
     0.9 &  0.4 & 0.01  & 0.00 &  0.60    & 0.79  & 5.11 & 0.78  & 13.51& 0.80 & 5.12 \\ \hline
     0.9 &  0.5 & 0.01  & 0.00 &  0.70    & 0.85  & 4.64 & 0.85  & 11.50 & 0.86 & 4.88 \\ \hline 
     0.9 &  0.6 & 0.01  & 0.00 &  0.79    & 0.90  & 3.82 & 0.91  & 08.18 & 0.92 & 3.90 \\ \hline

     0.9 &  0.0 & 0.001  & 0.35 & 0.50    & 0.81  & 3.17 & 0.81  & 7.17 & 0.81&3.18 \\ \hline 
     0.9 &  0.3 & 0.001  & 0.36 & 0.59    & 0.82  & 3.03 & 0.83  & 6.34 & 0.86&3.06 \\ \hline 
     0.9 &  0.4 & 0.001  & 0.35 & 0.67    & 0.87  & 2.89 & 0.87  & 6.18 & 0.89&3.03 \\ \hline
     0.9 &  0.5 & 0.001  & 0.36 & 0.75    & 0.90  & 2.89 & 0.89  & 5.85 & 0.93&2.96 \\ \hline
     0.9 &  0.6 & 0.001  & 0.36 & 0.83    & 0.94  & 2.74 & 0.93  & 5.33 & 0.96&2.79 \\ \hline

     0.99 &  0.0 & 0.01  & 0.00 & 0.08    & 0.89  & 17.43 & 0.90  & 56.20  & 0.89 &17.70 \\ \hline
     0.99 &  0.3 & 0.01  & 0.00 & 0.51    & 0.73  & 5.18 & 0.73  & 12.98 & 0.79 &5.77 \\ \hline
     0.99 &  0.4 & 0.01  & 0.00 & 0.62    & 0.81  & 5.09 & 0.82  & 13.01 & 0.84 &5.21 \\ \hline
     0.99 &  0.5 & 0.01  & 0.00 & 0.73    & 0.88  & 4.97 & 0.88  & 12.69 & 0.89 &5.04 \\ \hline
     0.99 &  0.6 & 0.01  & 0.00 & 0.81    & 0.92  & 3.66 & 0.92  & 7.35 & 0.93 &3.57 \\ \hline

     0.99 &  0.0 & 0.001  & 0.35 & 0.57   & 0.89  & 3.19 & 0.89  & 6.69 & 0.89&3.20 \\ \hline
     0.99 &  0.3 & 0.001  & 0.35 & 0.62   & 0.88  & 2.99 & 0.87  & 6.00 & 0.90&3.02 \\ \hline
     0.99 &  0.4 & 0.001  & 0.36 & 0.70   & 0.90  & 2.91 & 0.91  & 5.83 & 0.93&2.97 \\ \hline
     0.99 &  0.5 & 0.001  & 0.36 & 0.79   & 0.93  & 2.82 & 0.94  & 5.67 & 0.95&2.83 \\ \hline
     0.99 &  0.6 & 0.001  & 0.36 & 0.84   & 0.95  & 2.75 & 0.95  & 5.31 & 0.97&2.68 \\ \hline
     
  \hline    
     
 \end{tabular}
 
 \begin{tabular}{|c|c|c|c|c|c|c|c|c|c|c|c|c|}
 \hline
\multicolumn{3}{|c|}{other parameters} & \multicolumn{8}{|c|} {Simulation of different probabilities with $p_c= 0.99$} \\ \hline \hline
    $p_1$& $p_2$& $p_w$ & $S_1$ & $S_2$ &    $S_3$ & $N_3$& $S_4$ & $N_4$& $S_5$& $N_5$ \\ \hline
     0.9 &  0.0 & 0.01  & 0.00  & 0.08       & 0.90  & 18.2 & 0.89  & 55.62  & 0.90 &17.58 \\ \hline
     0.9 &  0.3 & 0.01  & 0.00  & 0.54       & 0.76  & 5.14 & 0.76  & 13.08 & 0.79 &5.64 \\ \hline
     0.9 &  0.4 & 0.01  & 0.00  & 0.67       & 0.85  & 5.05 & 0.85  & 12.92 & 0.87 &5.12 \\ \hline
     0.9 &  0.5 & 0.01  & 0.00  & 0.77       & 0.91  & 4.86 & 0.90  & 12.10 & 0.91 &5.02 \\ \hline
     0.9 &  0.6 & 0.01  & 0.00  & 0.86       & 0.94  & 3.57 & 0.93  & 7.24 & 0.95 &3.57 \\ \hline

     0.9 &  0.0 & 0.001  & 0.36  & 0.57      & 0.90  & 3.18 & 0.89  & 6.61 & 0.89 &3.19 \\ \hline
     0.9 &  0.3 & 0.001  & 0.36  & 0.65      & 0.88  & 2.98 & 0.89  & 6.34 & 0.92 &3.02 \\ \hline
     0.9 &  0.4 & 0.001  & 0.36  & 0.73      & 0.92  & 2.87 & 0.92  & 5.91 & 0.94 &2.91 \\ \hline
     0.9 &  0.5 & 0.001  & 0.35  & 0.81      & 0.94  & 2.81 & 0.94  & 5.51 & 0.96 &2.82 \\ \hline
     0.9 &  0.6 & 0.001  & 0.37  & 0.88      & 0.96  & 2.72 & 0.96  & 5.24 & 0.97 &2.90 \\ \hline

     0.99 &  0.0 & 0.01  & 0.00  & 0.09      & 0.98  & 16.9 & 0.98  & 51.40  & 0.98 &17.6\\ \hline
     0.99 &  0.3 & 0.01  & 0.00  & 0.58      & 0.83  & 5.66 & 0.83  & 14.73 & 0.87 &5.69 \\ \hline
     0.99 &  0.4 & 0.01  & 0.00  & 0.69      & 0.90  & 5.43 & 0.91  & 14.38 & 0.92 &5.69 \\ \hline
     0.99 &  0.5 & 0.01  & 0.00  & 0.80      & 0.94  & 4.61 & 0.94  & 11.19 & 0.95 &4.73 \\ \hline
     0.99 &  0.6 & 0.01  & 0.00  & 0.87      & 0.96  & 3.26 & 0.97  & 6.26 & 0.96 &3.35 \\ \hline

     0.99 &  0.0 & 0.001  & 0.35  &0.62      & 0.98  & 3.20 & 0.98  & 6.32 & 0.98  &3.28 \\ \hline
     0.99 &  0.3 & 0.001  & 0.36  &0.69      & 0.94  & 2.90 & 0.94  & 5.88 & 0.97  &3.00 \\ \hline
     0.99 &  0.4 & 0.001  & 0.36  &0.76      & 0.95  & 2.89 & 0.96  & 5.59 & 0.98  &2.85 \\ \hline
     0.99 &  0.5 & 0.001  & 0.36  &0.83      & 0.97  & 2.70 & 0.97  & 5.48 & 0.98  &2.80 \\ \hline
     0.99 &  0.6 & 0.001  & 0.36  &0.89      & 0.98  & 2.68 & 0.98  & 5.19 & 0.99  &2.69 \\ \hline

\end{tabular}
\end{center}
}

\label{simtable 1}
\end{table} 

Suppose $N_i$ denotes the average number of selected nodes to be searched corresponding to $S_i, \ i=1,2,\ldots,5$. Clearly, $N_1=1-S_1$ because 
we need no search when ${\cal M}_0$ is selected. When event occurs and we consider only one $N$ from ${\cal N}_{\max}$, we need at most
one search (since no search is needed if ${\cal M}_0$ is selected) and we have $N_2\leq 1$. 
In our simulation, we find $N_2=1$ in all the cases; that means, in simulation, ${\cal M}_0$ has not been selected
when event occurred. Note that $N_3 \geq 1$ since we consider all $N$'s in ${\cal N}_{\max}$ for searching. Again, as before, $N_4>N_3\geq1\geq N_2$. Also, by definition, $N_5\geq 1$.
Table \ref{simtable 1} presents the different $S_i$'s and $N_i$'s based on simulation for different values of $p_1,p_2,p_c$ and $p_w$ with $p_1$
and $p_c$ taking values 0.9 and 0.99, $p_w$ taking values 0.01 and 0.001 and $p_2$ taking values 0.0, 0.3, 0.4, 0.5 and 0.6. The choice of $p_1$
and $p_c$ reflects the corresponding high probability, whereas that of $p_w$ reflects small probability, which is desirable in a good sensor. Since the primary interest is to 
study the effect of detection by neighboring nodes, we consider $p_2$ as 0 (which means there is no effect of neighboring nodes) and some positive values less than $p_1$. 

Note that the probability of correct detection under ${\cal M}_0$ depends only on $p_w$. This is also evident in Table \ref{simtable 1}. Intuitively, if $p_w$ is high then the proportion $S_1$ of correct detection in normal situation is 
low. In Table \ref{simtable 1}, we see that $S_1$ is 0 for $p_w=0.01$, varies from 0.35 to 0.37 for $p_w=0.001$ and varies from 0.90 to 0.91 for $p_w=0.0001$ (not shown in Table \ref{simtable 1}). 
If we consider smaller value of $p_w$ then the success probability $S_1$ will be higher. Hence $p_w$ must be low as the number of hexagons is high to get better results in normal situation.

We see that the estimated false negative rate, that is an estimate of $P_{{\cal M}_N}({\cal M}_0 \mbox{~is selected})$, is often $0$ in our simulation (not shown
in Table \ref{simtable 1}). This is because, if the event occurs at $N$, then detection of the event by at least one of the nodes belonging
to $\{N\}\cup B(N)$ is highly probable. Furthermore, since the grid size is large, one of the node belonging to 
$R\setminus(\{N\} \cup B(N))$ may response
wrongly, though it cannot detect the event. So, under ${\cal M}_N$, there is a small probability to select ROI as normal.
If we take $p_w$ and the detection probabilities $p_1$ and $p_2$ to be very small, then we may get some positive false negative rate
but this is not a desired condition for a good sensor.

From simulation, we see that, as $p_2$ increases (for positive $p_2$), $S_i$ values increase whereas $N_i$ decrease.
As $p_2$ increases, it helps to differentiate between the likelihood values resulting in lower cardinality of the set ${\cal N}_{\max}$ and lower values of $N_i$'s.
However, since the neighboring nodes help to detect the event, the success probability increases. From simulation, we find that, as 
$p_1$ increases, success probabilities also increase, but the effect of $p_2$ is more prominent than that of $p_1$. On the other hand, 
success probabilities also change  with $p_w$ and $p_c$.
Since $p_2=0$ means $P_2=p_w$, so there is little variability in the likelihood values leading to larger size of ${\cal N}_{\max}$.
%To increase the success probability we consider a threshold value $C$ where $0<C<1$. If $L_{max}$ be the maximum likelihood among all the likelihoods, consider the set of nodes having likelihood greater than $C.L_{max}$ and we search whether the event node belongs to this set or not. 

When $p_w=0.01$, effect of $p_2$ on $S_3, S_4, S_5$ and $N_3, N_4, N_5$ seems to be significant, whereas 
the same cannot be said for $p_w=0.001$. There is sudden change in $S_i$'s and $N_i$'s, when we shift from $p_2=0$ to $p_2=0.3$, for 
$p_w=0.01$, but not  $p_w=0.001$. So, when $p_w$ is small, the effect of the neighborhood seems to be less.

%When $p_w=0.01$ $N_i$'s values are quite high and $S_i$'s are more for $p_2=0$ than that of $p_2=0.3$. But When $p_w=0.001$ $N_i$'s values are more and $S_i$'s are less for $p_2=0$ than that of $p_2=0.3$.

The values of $S_3$ and $S_4$ are very similar for different values of the parameters; but large increment in $N_4$ than $N_3$ suggests
that the idea of neighboring search is not effective. But $S_3$ is much higher than $S_2$; so the method of searching all the nodes in
${\cal N}_{\max}$ is a better idea than that of searching a random node from ${\cal N}_{\max}$.

We estimate the success probability $P_{\cal{M}_N}(L_N>C.L_{max})$ by simulation for different values of the threshold $C$ ranging from 0.5 to 0.9 (see Table~\ref{simtable 2}). 
Note that $S_5$ corresponds to the threshold value $C=0.9$. We consider $p_1=0.99,p_w=0.001,p_c=0.9$ and four values of $p_2=0.3, 0.4, 0.5, 0.6$. From Table~\ref{simtable 2}, we see that the success probability 
increases as the threshold value $C$ decreases and $p_2$ increases. Similarly, the number of search decreases with both $C$ and $p_2$.

\begin{table}[htb]
\scriptsize
\caption{Simulation of estimated success probabilities and number of searches for different threshold values ($C$) and some values of the parameters with $p_c= p_1=0.9$}
{
\begin{center}
\begin{tabular}{|c|c|c|c|c|c|c|c|c|c|c|c|c}
\hline
 \multicolumn{2}{|c|}{other parameters} & \multicolumn{2}{|c|}{$C=0.6$}&\multicolumn{2}{|c|}{$C=0.7$}&\multicolumn{2}{|c|}{$C=0.8$}&\multicolumn{2}{|c|}{$C=0.9$} \\ \hline \hline
     $p_2$& $p_w$& success & search     &success  & search  & success & search   & success& search  \\ \hline
     0.0 & 0.01  & 0.81 & 18.21    & 0.81  &18.25 & 0.81 & 18.21 & 0.81 & 18.17 \\ \hline
     0.3 & 0.01  & 0.87 & 13.72    & 0.78  & 9.13 & 0.75  &  6.64 & 0.73 & 5.70 \\ \hline 
     0.4 & 0.01  & 0.89 &  8.86    & 0.85  & 6.47 & 0.82  &  5.46 & 0.80 & 5.12 \\ \hline
     0.5 & 0.01  & 0.93 &  6.88    & 0.90  & 5.69 & 0.89  &  5.05 & 0.86 & 4.88 \\ \hline 
     0.6 & 0.01  & 0.97 &  5.27    & 0.96  & 4.91 & 0.93  &  4.04 & 0.92 & 3.90 \\ \hline 
 
     0.0 & 0.001  & 0.80 & 3.27    & 0.80  & 3.21 & 0.80  & 3.17 & 0.80&3.18 \\ \hline 
     0.3 & 0.001  & 0.91 & 4.15    & 0.91  & 3.65 & 0.87  & 3.31 & 0.86&3.06 \\ \hline 
     0.4 & 0.001  & 0.94 & 4.25    & 0.94  & 3.69 & 0.93  & 3.31 & 0.89&3.03 \\ \hline
     0.5 & 0.001  & 0.97 & 4.24    & 0.97  & 3.64 & 0.96  & 3.26 & 0.93&2.96 \\ \hline
     0.6 & 0.001  & 0.99 & 3.96    & 0.98  & 3.18 & 0.98  & 3.04 & 0.96&2.79 \\ \hline
     
  \hline    
     
 \end{tabular}
\end{center}
}

\label{simtable 2}
\end{table}

\section{Discussion}
One prime object of this paper is to show the effect of the neighboring nodes in detection of an event. In this section, we discuss 
the role of the neighboring nodes, some other related issues and make remarks.

\subsection{Role of the neighboring nodes}
Since $\ln L_N= a+ b(cz_N + t_N - dk(N))$, where $a, b,c $ and $d$ are as defined in Section \ref{Model}, $c$ denotes the weight of the 
central node compare to the neighboring nodes in the corresponding likelihood. Note that, since $P_1>P_2$, we have $c>1$ and, if $c$ is
close to 1, then the six neighboring nodes are as important
as the event node. So, as the value of $c$ increases, the importance of the neighboring nodes decreases. Also, $d$ gives some idea about the
role of the number of adjacent nodes, i.e., $k(N)$. Recall that $P_1$ and $P_2$ are the probabilities of responding (i.e., reporting the node $N$ as the
event hexagon) by the sensors $S_N$ and $S_{N'}$, respectively, when $N$ is the event hexagon and $N'$ is a neighboring node of $N$.
So, we numerically calculate the quantities $P_1, P_2, c$ and $d$ for some values of the parameters (see Table~\ref{simtable 4}).

From the theoretical results in Section \ref{Model} we see that, $ P_1 $ and $c$ increase as $p_1$ increases, while $P_2$ and $d$ do not depend on $p_1$.
On the other hand, while $ P_2 $ increases with $p_2$, $c$ and $d$ decreases and $P_1$ is independent of $p_2$. Therefore, the importance
of the neighboring nodes decreases with $p_1$ and increases with $p_2$, as expected and observed in Table~\ref{simtable 4}.

\begin{table} 
\scriptsize
\caption{Values of $P_1,P_2,c$ and $d$ for $p_c=0.9$}
{
\begin{center}
\begin{tabular}{|c|c|c|c|c|c|c|c|c|c|}
\hline
\multicolumn{2}{|c|}{parameters} & \multicolumn{4}{|c|} {$p_w=0.1$} & \multicolumn{4}{|c|} {$p_w=0.2$} \\ \hline \hline

    $p_1$ & $p_2$ &    $P_1$ & $P_2$ & $c$  & $d$   & $P_1$ & $P_2$ & $c$  & $d$ \\ \hline

    &    0.3 &     0.66 &  0.34 & 1.865 & 0.202 & 0.69 &  0.41 & 2.139 & 0.298 \\ 
       \cline{2-10}
     0.7    &    0.4 &     0.66 &  0.42 & 1.526 & 0.234 & 0.69 &  0.48 & 1.674 & 0.330 \\ \cline{2-10}
        &    0.5 &     0.66 &  0.50 & 1.302 & 0.268 & 0.69 &  0.55 & 1.378 & 0.363 \\ \cline{2-10} 
        &    0.6 &     0.66 &  0.58 & 1.135 & 0.302 & 0.69 &  0.62 & 1.166 & 0.397 \\ \cline{1-10}
 
     &    0.3 &     0.74 &  0.34 & 2.114 & 0.202 & 0.76 &  0.41 & 2.484 & 0.298 \\ \cline{2-10}
   0.8  &    0.4 &     0.74 &  0.42 & 1.730 & 0.234 & 0.76 &  0.48 & 1.944 & 0.330 \\ \cline{2-10}  
        &    0.5 &     0.74 &  0.50 & 1.476 & 0.268 & 0.76 &  0.55 & 1.600 & 0.363 \\ \cline{2-10}  
        &    0.6 &     0.74 &  0.58 & 1.287 & 0.302 & 0.76 &  0.62 & 1.354 & 0.397 \\ \cline{1-10}

       &    0.3 &     0.82 &  0.34 & 0.241 & 0.202 & 0.83 &  0.41 & 2.907 & 0.298 \\ \cline{2-10} 
    0.9 &    0.4 &     0.82 &  0.42 & 1.981 & 0.234 & 0.83 &  0.48 & 2.275 & 0.330 \\ \cline{2-10}  
        &    0.5 &     0.82 &  0.50 & 1.690 & 0.268 & 0.83 &  0.55 & 1.873 & 0.363 \\ \cline{2-10}  
        &    0.6 &     0.82 &  0.58 & 1.474 & 0.302 & 0.83 &  0.62 & 1.584 & 0.397 \\ \cline{1-10} 

\end{tabular}
\end{center}
}

\label{simtable 4}
\end{table}

\subsection{Estimation of the parameters}
In practice, the parameters $p_1, p_2, p_w$ and $p_c$ may be unknown. We can, however estimate the parameters by some experimentation. 

Note that, under ${\cal M}_0, Z_N$ follows $\Ber(p_w)$ for all $N\in R$. Hence, $p_w$ is the expected value of $Z_N$ given ${\cal M}_0$. So we perform the experiment by keeping the ROI normal. The proportion of $Z_N$'s having value $1$ gives an estimate of $p_w$. Repeat this experiment several times so that the average of the proportions over the repeated experiments can be taken as an estimate of $p_w$. 

Note that, $p_1$ is the expected value of $Y_N$ under ${\cal M}_N$. So, we perform the experiment by keeping an event in some node $N$ of the ROI. The proportion of $Y_N$'s having value $1$ gives an estimate of $p_1$. Repeat this experiment for several times so that the average of the proportions over the repeated experiments can be taken as an estimate of $p_1$. 
Similar experiments will give estimates of $p_2$ and $p_c$ as well.

\subsection{Incorporation of heterogeneity and uncertainty in parameters}
Let $\theta=(p_1,p_2,p_c,p_w)$ denote the set of parameters, which has been assumed to be the same for all the nodes. While, in practice there is no reason why the parameters should be same for all the nodes, it is also not clear how these would be different across $N$. This unexplained heterogeneity can be incorporated by assuming the $\theta$'s, for different $N$, to be independent realizations from a common distribution.

Let $\theta_N=(p_{1N}, p_{2N}, p_{cN}, p_{wN})$ denote the set of parameters for node $N$. We assume that $\theta_N, N\in R,$ are i.i.d. from some distribution, say, $g(\theta)$. Also assume that, given $\theta_N, \ N \in R, \ Z_N$'s are independent. Note that $g(\theta)$ denotes the joint distribution of the four parameters. For simplicity, we may assume them to be independent so that $g(\theta)$ can be written as $g(\theta)=g_1(p_1)g_2(p_2)g_c(p_c)g_w(p_w)$. In this situation, the likelihood for the model ${\cal M}_0$ is 
$$ \Pi_{N\in R} \int \ p_{wN}^{z_N}(1-p_{wN})^{(1-z_N)}g_w (p_{wN})dp_{wN},$$
where the integration is over the range of $p_{wN}$. Similarly, the likelihood for the model ${\cal M}_N$ can be written as 
$$\Pi_{N'\in R} \int L_N^{(N')}(\theta_N)g(\theta_N)d\theta_N,$$
where the integral is over the four-dimensional space given by the range of $\theta_N$, and $L_N^{(N')}(\theta_N)$ is the contribution of the $N'$th node to the likelihood $L_N$, given the value $\theta_N$, as described in Section \ref{Model}. 

Similar technique can also be used to incorporate parameter uncertainty. Even though the parameters can be assumed to be same for all the nodes, there may be reasonable uncertainty about the constancy of the parameter values. As in the Bayesian paradigm, the set of parameters may be assumed to be a realization from a distribution, say, $g(\theta)$. Then, the likelihoods for the model ${\cal M}_0$ and ${\cal M}_N$ are 
\begin{eqnarray*}
&&\int \Pi_{N\in R}\ p_{w}^{z_N}(1-p_{w})^{(1-z_N)}g_\theta (p_{w})dp_{w} ~~\mbox{and} \int \Pi_{N'\in R} L_N^{(N')}(\theta)g(\theta)d\theta ~~\mbox{respectively}.
\end{eqnarray*}
The choice of $g(\theta)$ may be a difficult one. However, sometimes there may be specific information available regarding the distribution of $\theta,$ which can be incorporated in the model. 

\subsection{When more sensors can detect the event square}
We may consider the situation when sensing radii are larger and more sensors can detect the event hexagon but with different 
probabilities. With respect to a particular node, classify the remaining nodes with respect to the probability of detecting the event at that node, which may as well depend on the distance from the particular node. Suppose that the sensors in the $i$-th class detect the event hexagon with probability $p_i, i=1,2,3,\ldots$. The theoretical analysis is similar to that of Section \ref{Analysis}, but having more probability terms.

\subsection{Concluding Remarks}
In this paper, we consider the problem of fault detection in wireless sensor network (WSN). We discuss how to address both the noise-related 
measurement errors ($p_1$ and $p_2$) and sensor fault ($p_c$ and $p_w$) simultaneously in fault detection, where the ROI is partitioned into
regular hexagons with the event occurring at only one hexagon. We propose fault detection schemes that explicitly introduce the error probabilities into the optimal event detection process. We develop the schemes under the consideration of model selection technique, multiple model selection technique and Bayesian model averaging method. The different error probabilities are calculated by means of simulation. Note that the same analysis can be carried out when ROI is partitioned into squares and sensors are placed at the centers. 

Nandi et al.~\cite{MBA} consider similar problem in wireless sensor network (WSN), in which the event can take place at the center of one
particular square (or hexagon) of the grid covering the ROI. In our paper, we allow the event square to be any one in the grid. 
Our approach can also be used for the problem of \cite{MBA} with only two models to be considered for selection. In~\cite{MBA}, the authors 
develop the scheme under the consideration of Neyman-Pearson hypothesis test, where the null and alternative hypotheses correspond to the two models. 
In model selection approach, we select the model with higher likelihood. In classical Neyman-Pearson hypothesis test, a model is selected if it's likelihood is greater than some constant times the likelihood of the other. This constant is fixed before the test depending on the size 
of the test. In model selection approach, the constant is $1$, leaving no choice for the size of the test. On the other hand, we cannot apply the classical Neyman-Pearson test with more than two models to be considered for selection. 

%Recall that, if $L_N<L_0$  for all $N$, then base station decide ROI as normal. Since detection of an event is more important than detection of ROI as normal, we can slightly modify are rule for base station as follows: decide $N$ as a event square if $L_N$ is maximum among all the likelihoods whether this value is greater than or less than or equal to $L_0$.

The principle of hypothesis testing places a large confidence in the null hypothesis and does not reject it unless there is strong evidence
against it. This safeguard of null hypothesis cannot be ensured in the model selection approach of Section \ref{Model}. However, the multiple model
selection approach of Section \ref{Multi} provides some safeguard in this regard.

This principle of model selection can be extended to the situation when there are more than one event hexagon and the objective is to detect the 
event hexagons. We may also assume that the sensors can detect different types of events. That is, response of sensors may not be only binary;
sensors can measure distance, direction, speed, humidity, wind speed, soil makeup, temperature, etc., and send the measurement of continuous type
variables to the base station. One needs a different formulation of the problem in such case which will be taken up in future.

\end{document}